\documentclass[12pt] {article}
\usepackage {psfig}

\newcommand {\bc}{\begin {center}}
\newcommand {\ec}{\end {center}}
\newcommand {\be}{\begin {equation}}
\newcommand {\ee}{\end {equation}}
\newcommand {\beq}{\begin {eqnarray}}
\newcommand {\eeq}{\end {eqnarray}}

\newcommand {\ovl}{\overline}
\def\disp {\displaystyle}
\def\intl {\int\limits}
\def\kB {k_{\rm B}}

\def\c {{\rm c}}
\def\d {{\rm d}}
\def\e {{\rm e}}
\def\m {{\rm max}}
\def\r {{\rm r}}
\def\B {{\rm B}}

\def\T {{\rm T}}
\def\X {{\rm X}}
\def\cI {{\cal I}}
\def\opt{{\rm opt}}
\begin {document}
 \hoffset-2.5cm \voffset-2.5cm

  \bc {\large\bf
THE SUNYAEV---ZEL'DOVICH EFFECT ON ELLIPTICAL GALAXIES} \ec
\bc{\bf \hspace{-1.3cm} B.\,V.\,Komberg$^1$, D.\,I.\,Nagirner$^2$,
I.\,V.\,Zhuravleva$^3$} \ec \bc{\it Astro space center of Lebedev
physical institute RAS, Moscow}$^1$

 {\it Sobolev astronomical institute of SPbSU, Saint-Petersburg}$^2$

 {\it Saint-Petersburg State University}$^3$\ec

 \bc ABSTRACT \ec

 \noindent The history of discovering of hot gas in galaxies is
traced briefly, its main properties are described and the
desirability to make them more precise, in particular to obtain
additional data on the mass of such gas is pointed out. For this
purpose observations of the Su\-nya\-ev---Zel'\-do\-vich effect on
hot gas of coronas of elliptic galaxies are proposed. The absolute
and relative disturbances of the cosmic microwave radiation
spectrum due to scattering of relic photons by Maxwellian
electrons are calculated according the formula of the article
\cite{ZeldSun}. With the example of three elliptic galaxies it is
shown that observation of the SZ effect on such galaxies is quite
possible. Kinematic SZ effect \cite{ZeldSunV} arising due to
peculiar movement and rotation of galaxies is available for
observation as well. Such observations combined with X-ray data
would make it possible to get more about properties of galactic
gas, to obtain additional information on rotation of galaxies, on
possible accreting gas flows and on hot galactic wind.

{\it Subject headings:} the Sunyaev---Zel'dovich effect, hot gas,
elliptical galaxies

\vspace{1cm}

\bc Introduction \ec

Effect of distortion of blackbody spectrum of microwave background
radiation (CBR) scattered by hot intergalactic gas in galaxy
clusters was predicted by Ya.B.Zel'dovich and R.A.Sunyaev in 1969
(known as SZ effect according to Latin alpha\-bet) and since then
is widely used  in the inves\-ti\-ga\-tions \cite{ZeldSun} of
properties of this gas and determination of some cosmological
parameters, \,for\, example\,, \,the Hub\-ble constant. Along with
thermal effect, ki\-ne\-ma\-ti\-cal effect was predicted
\cite{ZeldSunV} which is produced by the same scattering of relict
photons by gas with peculiar motions of clusters relative to CBR.
A lot of articles and surveys are devoted to extended description
and results of observations of both kinds of effect (for example,
\cite{ZeldSunRev}, \cite{SunZeldRev}, \cite{Rephaeli},
\cite{Birkrev}, \cite{Carlstrom}).

    Recent time SZ effect means not only scattering of CBR by hot gas
in cluster galaxies, but also scattering of CBR by other
astrophysical objects, for example by outflows of gas from active
galactic nuclei and  quasars (other examples are given in survey
\cite{Birkrev}). In this article we suggest to observe thermal and
kinematic effect on gas of el\-lip\-ti\-cal galaxies in this
article. Effect is about an order lower than on galaxy clusters
but nevertheless has access to observations. At the beginning of
the article we will briefly discuss a problem of hot gas in
galaxies.

\bc Hot Gas In Elliptical Galaxies  \ec

For a long time it was assumed that existence of interstellar gas
is a feature of disk galaxies, but not elliptical (E) or lens (S0)
galaxies, implying cold, neutral gas, the same as in our Galaxy.
Situation changed after launching X-ray satellites \cite{Fabiano},
\cite{evolk}. X-ray radiation of galaxy clusters (for example,
\cite{GiacKell}, \cite{Kell}, \cite{Tanan}, \cite{FabianAllen}),
which was discovered by satellite UHURU in the end of 1970, proved
the existence of hot gas in galaxies and intergalactic medium,
which radiates in X-ray. After increasing of angular resolution of
X-ray re\-ceivers a lot of point sources were discovered
\cite{Helfand}. Their contribution to X-ray radiation is important
to take into account \cite{Tucker}.

It became possible to study in details pro\-per\-ties of X-ray
ra\-diation from separate galaxies\cite{GiacBech}, including
subtraction of point sources ra\-dia\-tion and distinction of hot
diffuse gas radiation \cite{BierKron}, \cite{StanWar}  after
launching observatory Einstein in 1978. Hot gas was discovered in
five disk galaxies in Virgo cluster \cite{Forman} and than in
elliptical galaxies \cite{BuoteFab}, \cite{Buote}.

Later, using relations between X-ray $L_\X$ and optical $L_B$
galaxy luminosities which according to \cite{Buote},
\cite{CanizFabb} can be presented as $L_\X \sim
10^{19}(L_\B/L_\odot)^2$ {\it erg/s}, estimations of central
electron density  $\sim 0.1$ {\it cm $^{-3}$}, gas temperature
$\sim 10^7 ${\it K} and cooling time $\sim 5 \cdot 10^6 \div
5\cdot 10^7 $ years were deduced \cite{PaolisIngras}. By the way,
it was concluded that concentration of particles decreases with
increasing of distance $r$ from galaxy center according the power
law $\sim r^{-\alpha}$, where $\alpha \sim 1 \div 1.5$.

Relatively small cooling time in central regions of galaxies
suggests the existence of cooling flows. Dust clouds can remain in
cooling flows, infrared radiation of which in E/S0 galaxies is
observed by satellite IRAS \cite{Jura}, \cite{ThronsBal},
\cite{TemiBrigh}.  From this fact and the data of X-ray
observations \cite{BuoteFab} a conclusion can be made about
multiphase of interstellar medium not only in disk galaxies, but
in elliptical galaxies too. In connection with construction of
X-ray halo models and possible appearances of cooling flows such a
multiphase of ISM was considered, for example, in the article
\cite{Fabiano}. But, as it has been mentioned in series of papers,
for example in
 \cite{RosaGonzal}, star formation burst and activity of nucleus make a
contribution into ba\-lan\-ce between the heating and cooling of
gas.

Total mass of hot gas in massive elliptical ga\-la\-xies can reach
$10^{10} M_\odot$. According to \cite{BrighMath} the abundance of
heavy elements including iron increases from $Z_{Fe}\sim (0.2 \div
0.4)Z_\odot$ in periphery till $Z_{Fe}\sim (1 \div 2)Z_\odot$ in
the center. As it is shown in papers \cite{MatthiasGer},
\cite{Emsel} the amount of gas in galaxies and its X-ray
luminosity correlate with their stellar radiation hence with their
mass: $L_\X \sim L_{\opt}^2 \sim M_{stars}^2$. This conclusion is
true for galaxy clusters as well \cite{Komberg}.

A problem of estimation mass of virialized sys\-tems (they can be
E-galaxies and evoleved galaxy clusters) using properties of
diffused X-ray ra\-dia\-tion was discussed in several works
\cite{Vihlin}, \cite{Voevod}, \cite{Komberg}. It is clear that in
order to solve this problem it is necessary to know redistribution
of gas density in system along its radius and make some
assumptions about the degree of clustering in space, so the
solution of problem depends on model. Firstly, difficulty of
solution of this problem reveals in investigation of gas
components in galaxy clusters which in general are not virialized
systems. The evidences of this are observed heterogeneity in
surface brightness of diffuse X-ray radiation of hot gas in
galaxies \cite{Govani}, \cite{Rossetti}.

There are other methods of mass estimation of galaxies and galaxy
clusters. Firstly, it is a method based on the effect of
gravitational lensing \cite{Peng}, \cite{Gavazzi}. However, this
method needs a solution of re\-ver\-se problem namely recovering
of mass dis\-tri\-bu\-tion in gravitational lens by lensing image.
There is another way of mass estimation of ext\-ra\-ga\-la\-xy
systems connected with using in\-te\-gra\-ted thermal SZ effect,
offered by discovers of the effect \cite{ZeldSun}. This method is
less sensitive to heating and cooling of gas and does not depend
on the redshift of system. The arguments for using this method
instead of method based on estimation of mass of galaxy clusters
by their X-ray radiation are given in the work \cite{Motl}. These
advantages can appear in the case of investigation of hot coronas
of massive spheroid galaxies, vi\-riali\-za\-tion of which is
quicker and stops earlier, than in galaxy clusters. So such
coronas can be observed in X-ray on earlier cosmological times,
than galaxy clusters. Though the temperature of virialized
ga\-la\-xies coronas is an order lower, than in cluster gas, and
angular size is significantly less than the cluster size,
dis\-tri\-bu\-tion of gas in coronas is more ho\-mo\-ge\-neous,
because of quicker merge of galaxies from separate fragments in
comparison with clus\-ters. Observational facts give evidences of
this. Firstly, there are far quasars presenting active galactic
nucleus, when system of galaxies is far from vi\-ria\-lized state
and presented as protocluster. Secondly, there is dependence
between masses of central black holes and X-ray luminosities of
their coronas for massive elliptical galaxies, established in
\cite{MatthiasGer}, \cite{Emsel}, \cite{Komberg}.

As have been mentioned above, at the last years effect of
distortion of spectrum of CBR, due to scattering of radiation by
hot gas was used for investigation of series astrophysical
phenomena. For example, there is a discussion about cooling flows
in galaxy clusters in the work \cite{Majumdar}, about wind around
bright quasars in the article \cite{Natarajan}, \cite{Lapi}, about
young galaxies with powerful star formation burst in the work
\cite{RosaGonzal}.

In our article we pay attention to possibility of observation of
SZ effect on hot gas of spheroidal galaxies, which are not sources
of distant radio emission. Such galaxies used to be isolated
el\-lip\-ti\-cal galaxies, which were central in groups or not
rich clusters. \footnote {From this point of view, central
galaxies in rich clusters can not be used, because they are often
sources of strong radio emission with elongated components.
Besides, X-ray radiation of corona of central galaxy is hard to
separate from elongated radiation of cluster gas. According to
work \cite{unresolvradio} unresolved radio sources on 150 MHz can
introduce a distortion into SZ effect up to 10\%.} We consider
such observations to be interesting not only for physics of
galaxies, but for cosmology by the following reasons.

Firstly, distribution of gas in spheroid galaxies is more regular
and symmetrical, and is easier to study, than in clusters. Hence
estimations of gas distribution in simplified models are closer to
reality.

Secondly, X-ray luminosity function of coronas of galaxies is not
exposed such a strong "inverse evolution" as in galaxy clusters.
So effects con\-nec\-ted with existence of hot gas in galaxies can
be observed in large redshifts in comparison with clusters.

Thirdly, different correlations between pa\-ra\-me\-ters are
traced in elliptical galaxies, because of large degree of
virialization. Such parameters can be effective size by the
half-luminosity, effective brightness, velocity dispersion on
effective size, surface mass density or mass of central black hole
\cite{Bolton}. Similar correlations allow us to use such
ga\-la\-xies as "standard candle" or "standard length".

In the following part we will estimate the quan\-ti\-ty of effect
for three galaxies with known pa\-ra\-me\-ters and show that the
quantity is significant, so effect can be observed with modern
tools.

\vspace{1cm}

\bc Estimation of the Effect  \ec

1. {\it Mean optical depth of galaxy.} Let us es\-ti\-ma\-te a
possibility to observe effect by modern and ex\-pec\-ted tools.
Let us accept that galaxy and distribution of electron gas in it
has spherical symmetry, radius of galaxy $R$ and typical scale of
electron distribution $R_\c$. Let us draw a line through the
galaxy center from the observer and on distance $s$ from this line
and parallel to it the line of sight. Optical depth of galaxy by
Thompson scattering along the line of sight is defined by integral
over distance $y$ along the line of sight: \be \label{eq:tauex}
\tau_\e(s)=\sigma_\T\intl_{-\sqrt {R^2-s^2}}^{\sqrt {R^2-s^2}}
n_\e\left(\sqrt {s^2+y^2}\right)\d y, \ee where $\sigma_\T$ -
Thompson cross section , $n_\e$ - electron concentration on
distance $r$ from the center of galaxy.

We assume that electron concentration in ga\-la\-xy is described
by $\beta$-law distribution: \be \label{eq:ner} n_\e(r)=\frac
{n_\e^0}{\disp\left(1+r^2/R_\c^2\right)^\beta}, \ee where $n_e^0$
- electron concentration in galaxy center. Total amount of
electrons, corresponding to such decreasing electron density from
the center, is de\-fined as following \be \begin{array}{c}
\label{eq:Nebetarc} \disp N_\e=4\pi n_\e^0\intl_0^R\frac {r^2\d
r}{(1+r^2/R_c^2)^\beta}= \frac {4\pi}{3}n_\e^0
R^3\cI_\e(r_\c,\beta),\,\,\\\disp \cI_\e(r_\c,\beta)= \left(\frac
{r_\c^2}{1+r_\c^2}\right)^\beta F\left(\beta,1,\frac {5}{2}, \frac
{1}{1+r_\c^2}\right). \end{array} \ee Here $r_\c=R_\c/R$, $F(a, b,
c, x)$ - hypergeometric function. Dependence of integral
$I_\e(r_\c, \beta)$ of its arguments , which shows the difference
between total elec\-tron amount with $\beta$-law distribution with
pa\-ra\-me\-ter $r_\c$ and electron amount with homogeneous
dis\-tri\-bu\-tion in the whole volume of galaxy, is shown on the
figure 1.

\begin {figure}[ht]

\hspace {50mm} $\cI_\e(r_\c,\beta)$

\vspace {1mm}

\centerline {\psfig {file=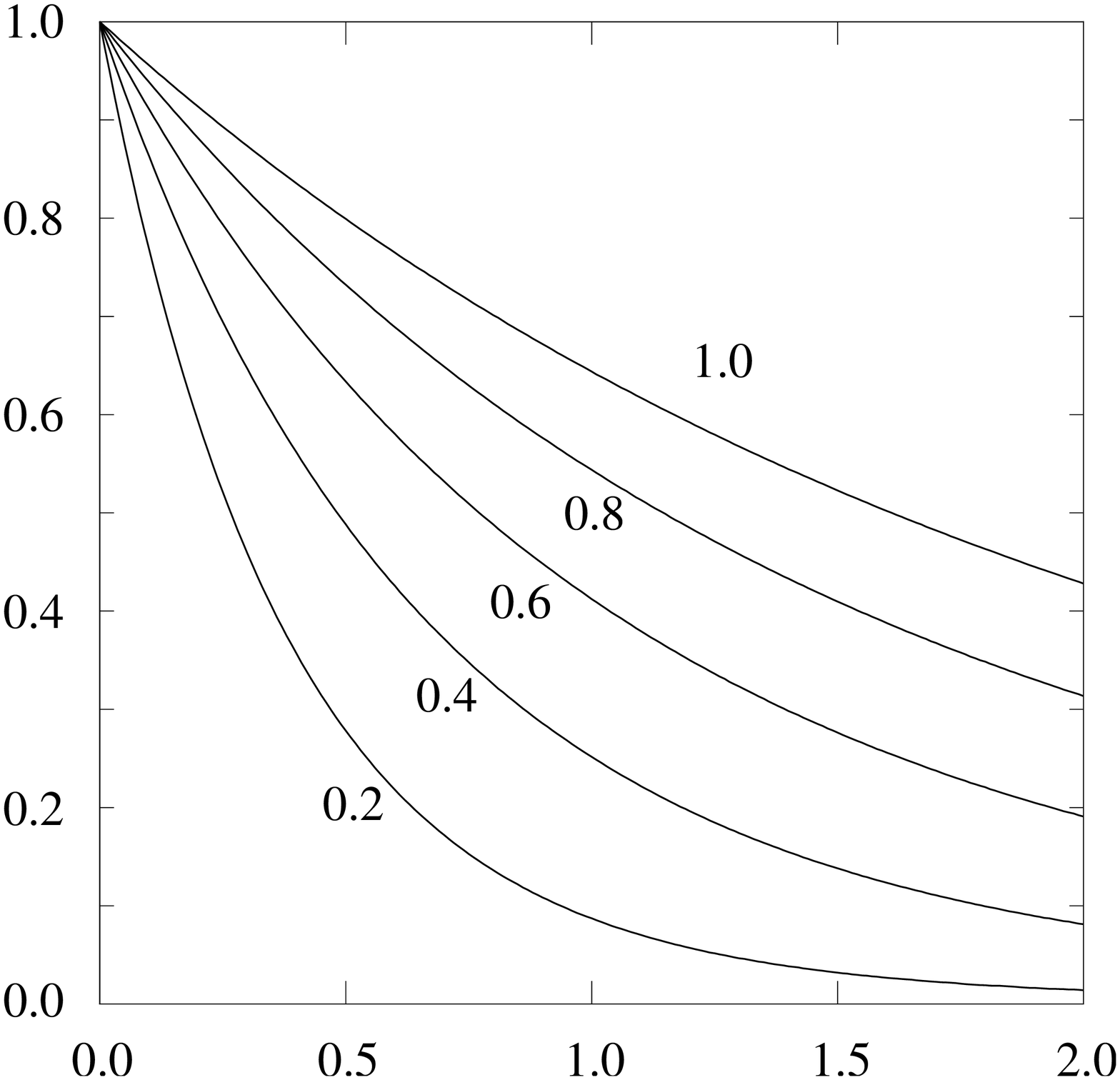,height=7cm}}

\vspace {-18mm} \hspace {70mm} $r_\c$

\vspace {3mm} \hspace {130mm} $\beta$

\vspace {7mm}

\centerline {Figure 1. Value of integral $I_\e(r_\c, \beta)$.}

\end {figure}

Optical depth in $\beta$-law distribution \be \label{eq:tauexne}
\tau_\e(s)\!=\!\sigma_\T n_\e^0\intl_{-\sqrt {R^2-s^2}}^{\sqrt
{R^2-s^2}} \frac {\d y}{\disp\left(1+\frac
{s^2+y^2}{R_\c^2}\right)^\beta}\!= 2\sigma_\T n_\e^0\sqrt
{R^2-s^2}\left(\frac {r_\c^2}{1+r_\c^2}\right)^\beta
F\left(\beta,1,\frac {3}{2},\frac {1-s^2/R^2}{1+r_\c^2}\right).
\ee

Averaging of optical depth of galaxy by $s$ leads to integral \be
\begin{array}{c}\label{eq:ovltaue} \disp \ovl {\tau}_\e\!=\!\frac
{2\pi}{\pi R^2}\intl_0^R\tau_\e(s)s\d s\!=\! \frac {4}{3}\sigma_\T
n_\e^0 R\cI_\tau(r_\c,\beta),\,\,\\\disp \cI_\tau(r_\c,\beta)\!=
\!3\left(\frac {r_\c^2}{1\!+\!r_\c^2}\right)^\beta\intl_0^1 u^2\d
u F\left(\beta,1,\frac {3}{2},\frac
{u^2}{1\!+\!r_\c^2}\right).\end{array} \ee

\begin {figure}[ht]

\hspace {50mm} $\cI_\tau(r_\c,\beta)$

\vspace {1mm}

\centerline {\psfig {file=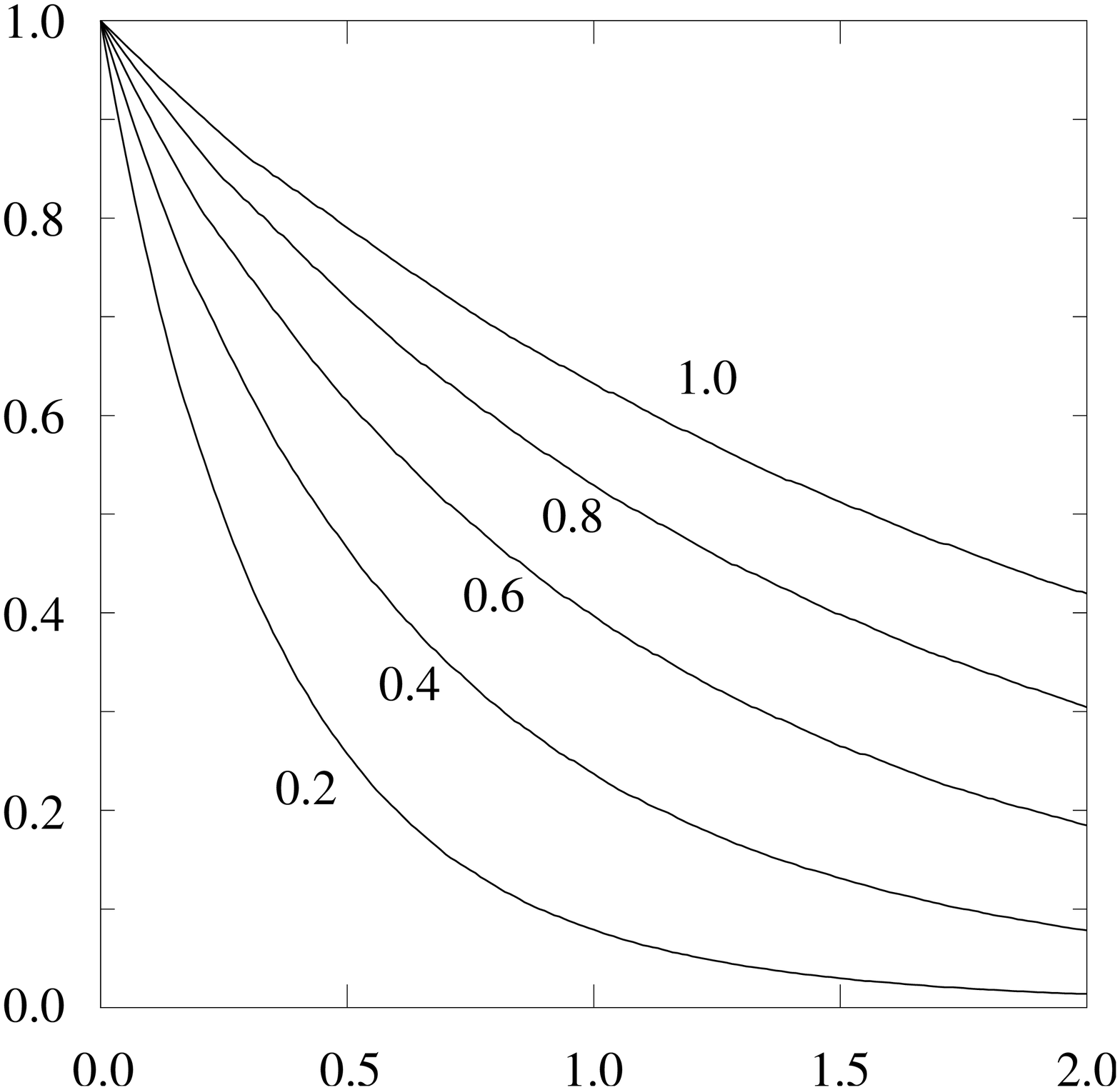,height=7cm}}

\vspace {-18mm} \hspace {70mm} $r_\c$

\vspace {10mm} \hspace {130mm} $\beta$

\centerline {Figure 2. Value of integral $\cI_\tau(r_\c,\beta)$.}

\end {figure}

Dependence of integral $I_\r(r_\c, \beta)$ of 0$\le \beta \le$ 2
are presented on the figure 2, where $r_\c=R_\c/R$ changes from
0.2 till 1.0 with step 0.2. Product $\disp{\tau
_\e^0=\frac{4}{3}\sigma_T n_\e^0R}$ is a maximum optical depth of
galaxy, i.e. depth taken along its diameter with homogeneous
electron distribution in galaxy($\beta=0$). Integral shows
difference of mean depth from maximum depth. Values of both
integrals are si\-mi\-lar, $I_\tau$ systematically lower than
$I_\e$, but maximum relative deviation of values $I_\tau$ from
va\-lues $I_\e$, presented on graphics, is equal 0.1 (with
$\beta$=1.1, $r_\c$=0.2).

2. {\it Influence of scattering on spectrum of CBR.} Parameters of
three galaxies, which are interesting for the calculation of
effect, are presented in table 1: electron temperature $T_\e$ in
{\it keV}, radius of region $R_\c$, luminous in X-ray, in {\it
kpc}, electron density in galaxy center $n_\e^0$ in {\it
cm$^{-3}$}, recession velocity $V_\r$ of galaxies in {\it km/s},
redshift $z$ and distance from galaxy $D$ in {\it Mpc}. The whole
data is taken from Internet \cite{galact}.

Let us estimate the changes of intensity of CBR, occurring because
of interaction of radiation with electron gas of galaxies. We will
use the for\-mu\-la for the SZ-effect, which was found by the
authors of the effect in the first approach, sup\-po\-sing that
gas in clusters is distributed ho\-mo\-ge\-neously \cite{ZeldSun}:
\be
\begin{array}{c}\label{eq:DeltaI} \disp \Delta I=\frac
{2h}{c^2}\left(\frac {\kB T_\r}{h}\right)^3 \frac {\ovl
{\tau}_\e}{y_\e}F(x_\r),\quad \\ \disp F(x)=\frac {x^4
e^x}{(e^x-1)^2} \left(\frac{x}{th(x/2)}-4\right).
\end{array}\ee\bc

{\bf Table 1.} Parameters of three elliptical galaxies

\medskip
{\footnotesize
\begin {tabular*}{8.3cm}{@{\extracolsep{\fill}}|r|c|c|c|c|c|c|}
\hline
Name & $T_\e$ & $R_\c$ & $n_\e^0$ & $V_\r$ & $z$ & $D$ \\
\hline
NGC\phantom {2} 499 & 0.9 & 89 & 0.1 & 4400 & 0.014673 & 73 \\
NGC 1332            & 4.3 & 48 & 0.1 & 1525 & 0,005084 & 26 \\
NGC 4291            & 4.3 & 61 & 0.1 & 1750 & 0,005861 & 29 \\
\hline
\end {tabular*}}
\ec

Here $x_\r = h\nu/(k_\B T_\r)$ - dimensionless frequency and
$T_\r=2.7 $ {\it K} - temperature of CBR. Dis\-tur\-ban\-ces
relative to Plank's function will be the following: \be
\label{eq:DeltaII} \frac {\Delta I}{I}=\frac {\ovl
{\tau}_\e}{y_\e}e^{x_\r/2} \frac {x_\r/2}{sh(x_\r/2)}\left(\frac
{x_\r}{th(x_\r/2)}-4\right). \ee

Usually distortion of Plank's spectrum is pre\-sen\-ted through
temperature distortion with the for\-mu\-la: \be \label{eq:DelTT}
\disp \frac {\Delta T}{T}=\frac {\Delta I}{I}\frac {\d\ln T}{\d\ln
I}= \frac {\Delta I}{I}\frac {1-e^{-x}}{x}, \ee so \be
\label{eq:DelTTr} \disp \frac {\Delta T}{T_\r}=\frac {\ovl
{\tau}_\e}{y_\e} \left(\frac {x_\r}{th(x_\r/2)}-4\right). \ee

We will use these formulas for elliptical galaxies.  Calculating
value of $\tau_\e^0$ and $\disp{y_\e=\frac{mc^2}{k_\B T_\e}}$, we
will accept that electron gas is isothermal and pa\-ra\-me\-ter
$R_\c=R$ (see table 2).

Logarithms of absolute values of intensity of CBR(in units of
spectral intensity, i.e. {\it erg/(c cm$^2$ Hrz)}) and logarithms
of relative disturbances  of temperature as a function of
wavelength in mm for three galaxies, \,characteristics \,of
\,which \,are shown in table 1, are presented on figures 3 and 4.
Changes of intensity are negative on wave length larger than
1.3775 cm, so logarithms are taken from their absolute values.

\begin {figure}[ht]

\hspace {53mm} $\disp\lg|\Delta
I|+\lg\frac{\tau_\e^0}{\ovl{\tau}_\e}$

\vspace {1mm}

\centerline {\psfig {file=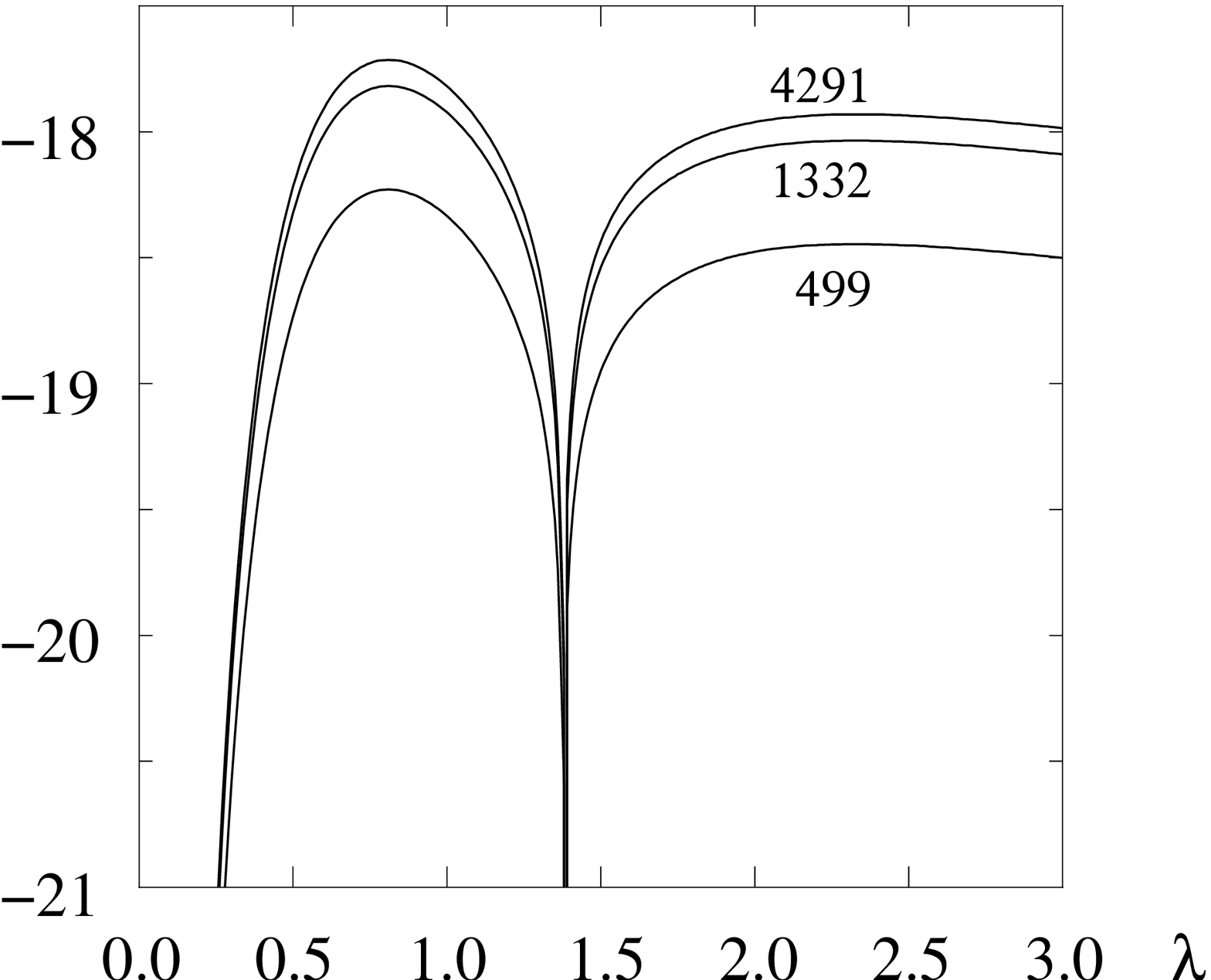,width=6cm}}

Figure 3. Absolute disturbances of intensity of CBR depending on
wave length in the directions to three galaxies.

\end {figure}

\begin {figure}[ht]

\hspace {53mm} $\disp\lg\left|\frac{\Delta T}{T_\r}\right|+
\lg\frac{\tau_\e^0}{\ovl{\tau}_\e}$

\vspace {1mm}

\centerline {\psfig {file=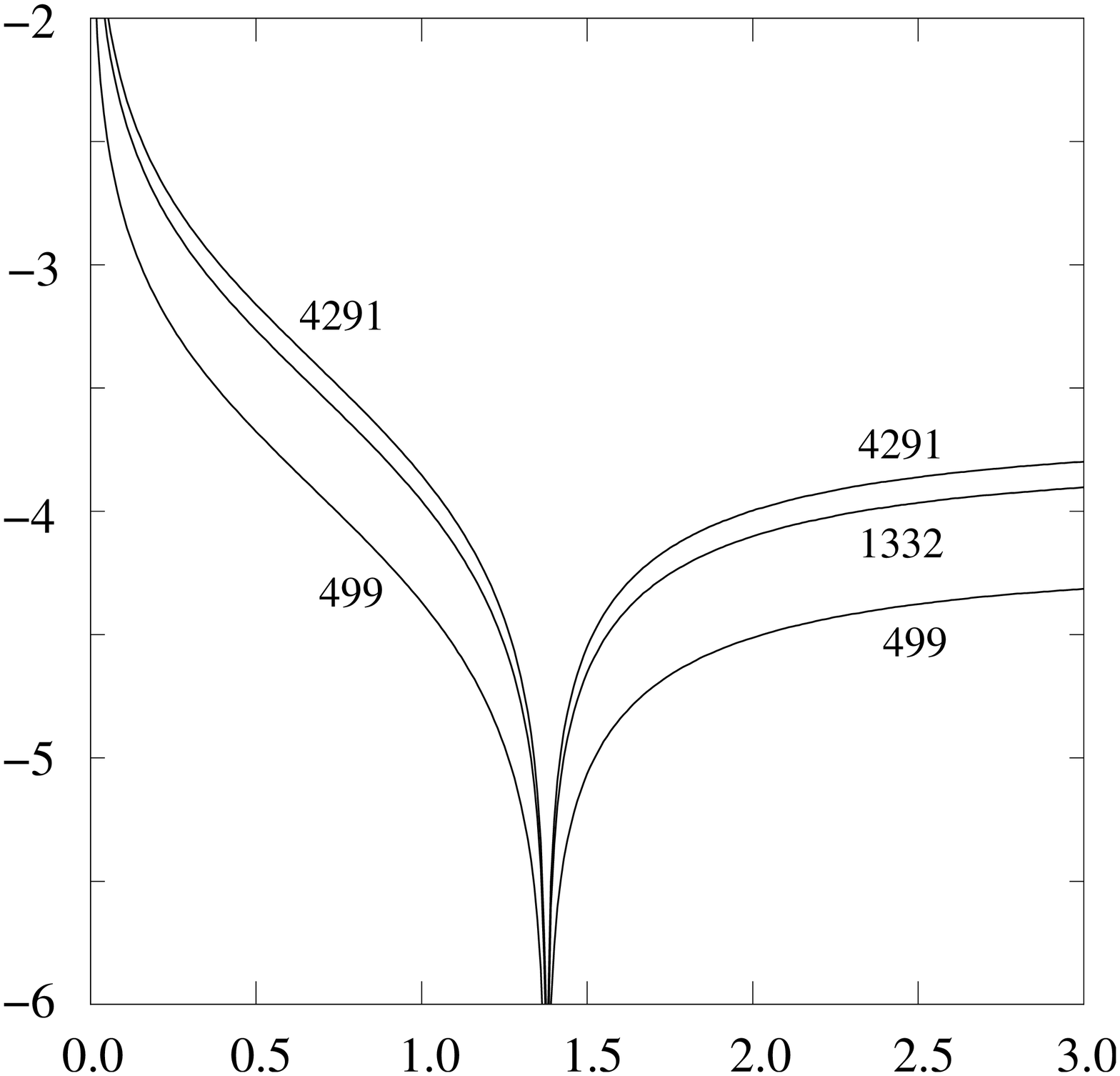,width=7cm}}

Figure 4. Relative disturbances of temperature of CBR depending on
wave length in the directions on three galaxies.

\end {figure}

Disturbances are equal to 0 at the point $\lambda_0=1.3775$ mm,
hence logarithm tend to $-\infty$. If $\lambda \to  0$,then
relative disturbances tend to $\infty$. We remind that relative
disturbances do not depend on $z$, because they are expressed only
through the func\-tion, depending on $x_\r$ - ration of frequency
and temperature, which both are proportional to $1+z$.

Disturbances depend on distance from the ga\-la\-xy center $r_\c$
so on the figures disturbances are presented and they are
calculated for maximum optical depths (therefore there are shown
addi\-tio\-nal term - logarithms of ratio of ma\-xi\-mum and mean
depth along the line of sight). Spectral disturbance of intensity
of CBR $\Delta I$ do not change, if we do not account for
cosmological expansion and absorption on the line of sight,
because tem\-pe\-ra\-tu\-re and frequency, used in ex\-pres\-sion
for dimensionless frequency are changed at the same way. Intensity
decreases as $(1+z)^2$, because of redshift and different temp of
time. This factor is nearly to unity, because redshifts of three
ga\-la\-xies are not large. Intensity is calculated on the unity
of space angle, but receiver fixes radiation in definite solid
angle, which is equal to square of galaxy divided on squared
distance till galaxy: $\Delta \Omega = \pi R_\c^2/D^2$. Angular
sizes of galaxies in arcminutes $\Delta\theta=(180\cdot
60/\pi)R/D$ and solid angles of galaxies, spectral fluxes of CBR
$H_{max}= I_{max}\Delta \Omega$ in maximum dis\-tur\-ban\-ce of
CBR (on $\lambda_{max} = 0.81$ mm), additional fluxes, because of
scattering $\Delta H_{max} = \Delta I_{max}\Delta \Omega$ in {\it
Jy} are presented in table 2 too. Parts of additional fluxes and
relative temperature perturbations are shown in table at the same
wave length.

Here are estimations for wave length 0.81 mm, which belongs to
Wien spectral region of CBR, where scattering encreases intensity.
The largest part of observations of SZ belongs to Rayleigh-Jeans
region, in which intensity decreases. It is seen from figures 3
and 4 that in Wien region effect strongly depends on $\lambda$, at
$\lambda >1.38$ mm this dependence is more smooth. For example, to
get $\Delta H_\m/H_\m$ and $(\Delta T/T_\r)_\m$ at $\lambda=3$ mm,
we have to multiply quantities from table 2 multiply on 0.596, and
at $\lambda=10$ mm - on 0.771.

We will remind that presented quantities are calculated for
maximum optical depths of electron gas of galaxies. For the
solution of integral characteristics we have to multiply them on
$\ovl{\tau}_\e/\tau_\e^0$, which depend on  galaxy parameters
$r_\c$ and $\beta$ and are shown in figure 2. This will decrease
the value of effect. For example, for $\beta$=1 and $r_c$=0.2,
0.4, 0.6, 0.8, 2.0 corresponding factors are 0.078, 0.237, 0.370,
0.529, 0.632.

Along with scattering on thermal electrons, macroscopic motions
influence on spectrum of CBR: peculiar motion of galaxies and
rotation. This additional kinematical SZ-effect leads to the
following distortion of spectra \cite{ZeldSunV}: \be
\label{eq:DeltaIkin} \disp \Delta I=-2\frac {(\kB
T_\r)^3}{h^2c^2}\frac {v_\r}{c}G(x_\r),\quad G(x)=\frac {x^4
e^x}{(e^x-1)^2}, \ee where $V_\r$ - radial velocity of electron
gas relative to CBR.

We can consider peculiar velocities of galaxies to be not lower
than velocities of motion of ga\-la\-xy clusters, so kinematic
effect also can be observed. Calculated values of disturbances of
CBR and consideration about their peculiar velocities allow \bc

Table 2. Distortion of spectrum of CBR for three elliptical
galaxies

\medskip

\begin {tabular*}{17.3cm}{@{\extracolsep{\fill}}|r|c|c|r|r|r|c|c|c|}
\hline Name & $y_\e$ & $\tau_\e^0$ & $\Delta\theta$ &
$\Delta\Omega\phantom{aaa}$
& $H_\m$ & $\Delta H_\m$ & $\Delta H_\m/H_\m$ & $(\Delta T/T_\r)_\m$ \\
\hline NGC\phantom {2} 499 & 568 & $1.83\cdot 10^{-2}$ & 4.19 &
$47\cdot 10^{-7}$
 &  522 & 0.28 & $5.4\cdot 10^{-4}$ & $8.14\cdot 10^{-5}$ \\
NGC 1332            & 119 & $0.99\cdot 10^{-2}$ & 6.36 & $107\cdot
10^{-7}$
 & 1188 & 1.63 & $1.4\cdot 10^{-3}$ & $2.10\cdot 10^{-4}$ \\
NGC 4291            & 119 & $1.25\cdot 10^{-2}$ & 7.25 & $140\cdot
10^{-7}$
 & 1554 & 2.70 & $1.7\cdot 10^{-3}$ & $2.67\cdot 10^{-4}$ \\
\hline
\end {tabular*}
\ec us to consider that observations of SZ-effect on electron gas
in elliptical galaxies are possible with modern tools. Indeed,
anisotropy of CBR is ob\-ser\-ved on level $10^{-5} \div 10^{-6}$
relative to background \cite{Rephaeli}, which exceeds necessary
quantity according to figure 4. The same conclusion can be done
for absolute values of distortions, because sensitivity of
antennas of modern radio telescopes in mil\-li\-me\-ter and
submillimeter range is about 1 {\it Jy} in second \cite{apparats},
 hence effect can be ob\-ser\-ved for the more distant galaxies too.
Sa\-tel\-li\-te PLANCK, whose launch is planed in 2008, allows to
increase this distances.

\bc Conclusions \ec

We showed that elliptical galaxies are objects, which as galaxy
clusters can noticeably distort spectrum of CBR by scattering in
hot gas. In spite of the temperature of gas is lower, than in
clusters, and galaxies are observed on shorter distances\,,
\,\,their \,\,properties \,\,are \,known \,much better, than
properties of clusters. Par\-ti\-cu\-lar\-ly, we can take into
account the influence of rotation, unhomogeneous of gas
distribution, its temperature and also influence of hot gas flows
of different nature. According to \cite{Birkrev} the value of
thermal effect on galaxy clusters varies from $-3.5\cdot 10^{-3}$
K till $6\cdot 10^{-3}$ K, but their are clusters for which the
effect is an order higher. In the case of elliptical galaxies the
value of effect in Rayleigh-Jeans region is an order lower, for
considered galaxies it correspondes to $5\cdot10^{-5}$ K,
$1.25\cdot 10^{-4}$ K and $1.59\cdot 10^{-4}$ at $\lambda=3$ mm.

The case of scattering CBR by hot gas, outflowing from AGN is not
less interesting. Re\-cent\-ly a work of E. Scannapieco et al
\cite{Scan} has ap\-peared, in which the influence of hot gas from
AGN on small scale distribution CBR is considered and dependence
of temperature distortion of CBR from redshift of galaxies is
obtained, both thermal and kinematical SZ-effect being considered.
In spite of the effect was examined in AGN (not in elliptical
galaxies), results are in good accordance. Thus, observations of
SZ-effect in elliptical galaxies and AGN (along with
ob\-ser\-va\-tions of these galaxies in other wavebands) can give
us additional information about their structure and properties of
components, also they will allow us to make more precise parameter
of cosmological models.

 \end {document}